\documentclass{aa}  

\usepackage{graphicx}
\usepackage{txfonts}
\usepackage{natbib,twoopt}
\usepackage[breaklinks=true]{hyperref}
\usepackage{multirow}
\bibpunct{(}{)}{;}{a}{}{,} 
\usepackage{xcolor}
\usepackage{arydshln}
\begin{document}

\title{(121514) 1999 UJ$_{7}$: A primitive, slow-rotating Martian Trojan\thanks{Based on service
    observations made with the 4.2 m William Herschel Telescope operated on the island of La Palma
    by the Isaac Newton Group of Telescopes in the Spanish Observatorio del Roque de los Muchachos
    of the Instituto de Astrof\'{i}sica de Canarias and on data collected with 2 m Ritchey-Chr\'{e}tien-Coud\'{e} (2mRCC) Telescope
    at Rozhen National Astronomical Observatory.}}


   \author{G. Borisov\inst{1,2}
          \and
          A. A. Christou\inst{1}
          \and
          F. Colas\inst{3}
          \and
          S. Bagnulo\inst{1}
          \and
          A. Cellino\inst{4}
          \and
          A. Dell'Oro\inst{5}
          }

   \institute{Armagh Observatory and Planetarium, College Hill, Armagh BT61 9DG, Northern Ireland, United Kingdom\\
              \email{Galin.Borisov@Armagh.ac.uk}
         \and
         Institute of Astronomy and NAO, Bulgarian Academy of Sciences, 72, Tsarigradsko Chauss\'ee
         Blvd., BG-1784 Sofia, Bulgaria
         \and
         IMCCE, Observatoire de Paris, UPMC, CNRS UMR8028, 77 Av. Denfert-Rochereau, 75014
         Paris, France
         \and
         INAF - Osservatorio Astrofisico di Torino, via Osservatorio 20, 10025 Pino Torinese
         (TO), Italy
         \and
         INAF - Osservatorio Astrofisico di Arcetri, Largo E. Fermi 5, I-50125, Firenze, Italy
             }

   \date{Accepted XXX. Received YYY}

 
  \abstract
 {}
 {The goal of this investigation is to determine the origin and surface composition of the
   asteroid (121514) 1999 $\mbox{UJ}_{7}$, the only currently known $\mbox{L}_{4}$ Martian Trojan
   asteroid.}
 {We have obtained visible reflectance spectra and photometry of 1999 $\mbox{UJ}_{7}$ and compared the spectroscopic
   results with the spectra of a number of taxonomic classes and subclasses. A
   light curve was obtained and analysed to determine the asteroid spin state.}
 {The visible spectrum  of 1999 $\mbox{UJ}_{7}$ exhibits a negative slope in the
   blue region and the presence of a wide and deep absorption feature centred around
   $\sim$0.65\,$\mu$m. The overall morphology of the spectrum seems to suggest a C-complex taxonomy.
   The photometric behaviour is fairly complex. The light curve
   shows a primary period of \mbox{1.936\,d}, but this is derived using only a subset of the 
   photometric data.
   The asteroid may be in a non-principal axis rotational state, but our observational coverage is
   insufficient to draw definitive conclusions.}
 {Although the observed spectral absorption is wider and deeper, this finding may be compatible with the 0.7\,$\mu$m
   spectral feature exhibited by some \mbox{Ch-type} asteroids and could possibly be interpreted as diagnostic of
   the presence of hydrated minerals. The inferred composition of 1999 $\mbox{UJ}_{7}$ as a primitive
   object can be consistent with a volatile-rich object originally accreted beyond the snow line of the solar system, and subsequently evolved to reach the inner regions of the solar system.}
 

   \maketitle
%

\section{Introduction}

The so-called Mars Trojans are asteroids orbiting in the stability regions corresponding to the
$\mbox{L}_{4}$ and $\mbox{L}_{5}$ Lagrangian points of Mars. 
Their existence is thought to date back to early epochs of the history of the solar system
\citep{Scholl.et.al2005}.
Among the nine confirmed Martian Trojans (as of 2017), eight belong to $\mbox{L}_{5}$ and only one, which is the
subject of the present study, belongs to $\mbox{L}_{4}$.

Seven of the $\mbox{L}_{5}$ Mars Trojans are found to be members of the Eureka family that were 
first identified by \citet{Christou2013} and \citet{deLaFuenteMarcoses2013}. 
Recently, \citet{BorisovMarTr2017} and \citet{Polishook.et.al2017} found that the surface composition of
the known Eureka family members seems to be dominated by olivine. This led
\citeauthor{Polishook.et.al2017} to suggest that these objects might be pieces of the Martian mantle,
excavated from the planet's interior by a giant impact and deposited in its Trojan 
clouds during the late phases of terrestrial planet formation. This is currently only a speculative
possibility, but in any case the inferred composition of these $\mbox{L}_{5}$ objects seems to
correspond to what should be expected for bodies having experienced an important thermal evolution
\citep{Rivkin.et.al2003, Rivkin.et.al2007,BorisovMarTr2017}.
The situation may be very different in the case of asteroid (121514) 1999 $\mbox{UJ}_{7}$ (``UJ7''
hereafter). The composition of this asteroid, the only $\mbox{L}_{4}$ Mars Trojan known, may have
important implications for current models of early solar system evolution, as explained below.
During a spectroscopic survey of Mars Trojans, \citet{Rivkin.et.al2003} obtained a visible
$\lambda\in$[0.55-1.00]\,$\mu$m spectrum of UJ7. They interpreted the smooth, featureless spectrum as
possibly diagnostic of an object belonging to the X taxonomic class. As noted by these authors,
the X-class, according to both \citet{TholenPhD} and \citet{BusBinzel2002}
classifications, encompasses a large variety of objects, including possible parent bodies of enstatite 
meteorites and iron meteorites, as well as organic-rich primitive meteorites. 
In the Tholen classification, the X-class is ambiguous and 
encapsulates three separate subclasses: E, M, and  P. Uniquely assigning 
any one of these taxonomies to a spectrum requires independent knowledge of the albedo. 
The Tholen P-class was found in the
80s to include low-albedo objects mostly orbiting in the outer main belt and in the Jupiter Trojan region
\citep{Emery.et.al2011}, generally thought to have primitive, volatile-rich compositions.
The low visible albedo ($p_V=0.053\pm0.034$) obtained by the NEOWISE survey \citep{Nugent.et.al2015} might
in principle be used to constrain the taxonomic classification and interpretation, but it is too
uncertain to draw definitive conclusions. Interestingly enough, \citet{Rivkin.et.al2003} noted that the
visible spectrum of UJ7 was also found to be similar, within observational uncertainties, to that of
(308) Polyxo, an asteroid belonging to the T-class defined by \citet{BusBinzel2002}. The T-class is also
very heterogeneous in terms of possible meteorite analogues. It has been interpreted
\citep{Britt.et.al1992} as diagnostic of a troilite-rich composition, suggesting important thermal
processing, because the troilite is a mineral found in mesosiderite meteorites and thought to be present in
the interiors of differentiated asteroids. An alternative interpretation of the T-class is that the body surface 
may be composed of altered carbonaceous chondrite material \citep{Bell.et.al1989}. It is 
worth noting that the T-class is also considered to be a good spectral match for the leading
side of the Martian satellite Phobos \citep{Rivkin.et.al2002}.

Of course, the mutually exclusive hypotheses that UJ7 could be either a thermally evolved body or a
primitive object depict very different scenarios. In the former case, the properties of UJ7 would be
consistent with the interpretation of Martian Trojans as surviving samples of an early population of
differentiated planetesimals present in the inner solar system at the epoch of the formation of Mars
\citep{Sanchez.et.al2014}. This would also provide a direct compositional link between Mars Trojans
and the material in the Martian satellite system (Phobos).
In contrast, a primitive composition of UJ7 would rather suggest genetic links with objects accreted in
the outer solar system. This would be compatible with the hypothesis of an early inward incursion of
material from the outer solar system, as predicted by recent models of early planetary evolution
\citep{Walsh.et.al2011, Vokrouhlicky.et.al2016}.

The limited spectral coverage and low signal-to-noise ratio (S/N) of the \citet{Rivkin.et.al2003} spectrum was not sufficient to
draw robust conclusions about the most likely composition of UJ7.
This motivated us to carry out new spectroscopic and photometric observations of this object to improve
our understanding of its properties and obtain some more firm evidence about its most likely
origin. 

The paper is organised as follows: in Section~\ref{obs} we present our observations. The 
data reduction procedures are described in Section~\ref{datared}, and the results in Section~\ref{results}, 
separately for spectroscopy and photometry. A discussion of our results is given in 
Section~\ref{blabla}.

\section{Observations}
\label{obs}
We obtained spectra of UJ7 using the Auxiliary port CAMera (ACAM) mounted at the folded-Cassegrain focus of
the 4.2 m William Herschel Telescope (WHT), Isaac Newton Group (ING) at the Roque de los Muchachos
Observatory, Canary Islands on 17 Feb 2017 and using the 2-channel focal reducer (FoReRo2) of the
2 m Ritchey-Chr\'{e}tien-Coud\'{e} (2mRCC) telescope at the Bulgarian National Astronomical Observatory
(BNAO), Rozhen on 25 Dec 2016.
Photometric observations were carried out using both FoReRo2 and the CCD camera at 
the 1 m telescope (T1m) at the Observatoire Midi-Pyr\'{e}n\'{e}es (OMP) 
situated on the technical platform of the Pic du Midi observatory, on the
following six nights: 1, 2, 5, 6, and 29 Dec 2016 and 1 Jan 2017, when the asteroid was at heliocentric distance r=1.48\,au and geocentric distance $\Delta$=0.63\,au and its phase angle changes from 30 to 27 degrees.
A Sloan $\mbox{r}^\prime$ filter was used.
Details of our observations are shown in Tables~\ref{specobs} and~\ref{photobs}.

\subsection{Spectroscopy}
\subsubsection{ACAM}
The low-resolution spectroscopy mode of ACAM was used for high-throughput
low-resolution spectroscopy in the range 3500--9400\,\AA\,. On-axis resolution at 6000\,\AA\,is R$\sim$450
and 900 for 1.0- and 0.5-arcsec slits, respectively. We used a GG495 order-blocking filter to obtain a
region uncontaminated by light from other orders from 4950
to 9500\,\AA.\footnote{Details of ACAM spectroscopy are available at the ING website:
  http://www.ing.iac.es/Astronomy/instruments/acam/spectroscopy.html}. The solar analogue  HD\,28099 was
observed to calibrate the asteroid spectrum.

\subsubsection{FoReRo2}
The instrument adapts the imaging elements of the detector to the characteristic
size of the object or the seeing disc. Built mainly for observations of cometary plasma, FoReRo2 has
proven to be suitable for many other tasks \citep{FoReRo2}. 
This instrument allows long-slit spectroscopy with a Bausch \& Lomb grating prism, working in the parallel beam of
FoReRo2, with 300\,lines/mm giving a spectral resolution of 4.3\,\AA/px using a 200\,$\mu$m slit width or
2.6\,arcsec on the sky.
The solar analogue HD\,86728 was observed to calibrate the asteroid spectrum.
\begin{table}
\setlength{\tabcolsep}{1pt}
\centering
\caption[Spectroscopic observations.]{Spectroscopic observations.}
\label{specobs}
\begin{tabular}{rcccccrr}
\hline 
\hline
\multirow{2}{*}{Object} & \multirow{2}{*}{Date} & \multirow{2}{*}{UT} &  Air-    & App.    & Exp.T & \multirow{2}{*}{Instrument}\\
                                    &                                   &                               &  mass & Vmag.    & N$\times$(sec) & \\
\hline
\hline
UJ7 & 17 Feb 2017 & 02:00 & 2.00 & 19.04   & 2$\times$600  & \multirow{2}{*}{ACAM} \\
HD\,28099 & 17 Feb 2017 & 00:20 & 2.13 & 8.12   & 1$\times$5 & \\
\hline
UJ7 & 25 Dec 2016 & 03:05 & 1.49 & 18.11   & 2$\times$600 & \multirow{2}{*}{FoReRo2}\\
HD\,86728 & 25 Dec 2016 & 03:54 & 1.09 & 5.40   & 1$\times$2 & \\
\hline
\end{tabular}
\end{table}

\subsection{Sloan $\mbox{r}^\prime$ photometry}
\subsubsection{FoReRo2} 
The imaging mode of the instrument was used for photometry of UJ7. 
For this investigation we used a Sloan $\mbox{r}^\prime$ filter, therefore only the red
channel of the instrument was in operation.
\subsubsection{T1m} 
The observations were performed with a DZ936BV Marconi CCD, using the same Sloan $\mbox{r}^\prime$ filter.
\begin{table}
\setlength{\tabcolsep}{4pt}
\centering
\caption[Photometric observations.]{Photometric observations.}
\label{photobs}
\begin{tabular}{cccccrc}
\hline 
\hline
 \multirow{2}{*}{Date} & \multirow{2}{*}{UT} &  Air-    & App.    & \multirow{2}{*}{Flt.} & Exp.T & \multirow{2}{*}{Instr.}\\
                                   &                               &  mass & Vmag.   &                                   & N$\times$(sec)  & \\
\hline
\hline
 05 Dec 2016 & 16:50   & 1.946 &               &                        &                              & \parbox[t]{2mm}{\multirow{9}{*}{\rotatebox[origin=c]{90}{FoReRo2}}}\\
     \textbar     & \textbar& 1.101 & 18.20 & $\mbox{r}^\prime$ & 295$\times$120  & \\
 06 Dec 2016 & 04:35   & 1.384 &               &                        &                               & \\
\cdashline{1-6}
 06 Dec 2016 & 18:20   & 1.565 &               &                        &                             & \\
     \textbar     & \textbar& 1.102 & 18.19 & $\mbox{r}^\prime$ & 255$\times$120   & \\
 07 Dec 2016 & 04:35   & 1.398 &               &                        &                            &  \\
 \cdashline{1-6}
 01 Jan 2017 & 19:00   & 1.137 &               &                        &                           &   \\
     \textbar    & \textbar& 1.085 & 18.15 & $\mbox{r}^\prime$ & 240$\times$120 &   \\
 02 Jan 2017 & 04:40   & 2.106 &               &                        &                            &   \\
\hline
 01 Dec 2016 & 21:30   & 1.361 &               &                        &                           &   \parbox[t]{2mm}{\multirow{9}{*}{\rotatebox[origin=c]{90}{CCD@T1m}}}\\
     \textbar     & \textbar& 1.085 & 18.20 & $\mbox{r}^\prime$ & 340$\times$90 &   \\
 02 Dec 2016 & 06:15   & 1.325 &               &                        &                            &   \\
 \cdashline{1-6}
 02 Dec 2016 & 19:30   & 1.699 &               &                        &                           &   \\
     \textbar     & \textbar& 1.086 & 18.19 & $\mbox{r}^\prime$ & 390$\times$90 &   \\
 03 Dec 2016 & 05:30   & 1.252 &               &                        &                            &  \\
 \cdashline{1-6}
 29 Dec 2016 & 17:55   & 1.420 &               &                        &                           &   \\
  \textbar    & \textbar & 1.081 & 18.15 & $\mbox{r}^\prime$ & 235$\times$90 &   \\
30 Dec 2016 & 00:25   & 1.113 &               &                        &                            &   \\
\hline
\end{tabular}
\end{table}

\section{Data reduction}
\label{datared}
\subsection{Spectroscopy -- ACAM and FoReRo2} 
All spectra were reduced under the assumption of
point-like sources and had their instrument signature removed by de-biasing, flat-fielding,
wavelength calibration, signal extraction, and sky subtraction with the
\textsc{iraf} packages \textsc{ccdred} and \textsc{onedspec}.

To maximise the S/N, the reduced 1D spectra for both the asteroid and solar analogue
stars were rebinned in 7.5\,nm steps, which is 22 times coarser than the spectrograph resolution
($\sim$0.34\,nm/px), after applying a $\sigma$-clipping cleaning algorithm in the same wavelength
window. The asteroid spectrum was then divided by the solar analogue spectrum and the result
was normalised to unity at $\lambda=550$\,nm. 
For FoReRo2 we used different rebinning steps in the blue and red parts of the spectra at 7.5\,nm and
30\,nm, respectively, to account for the lower S/N ratio at longer wavelengths. 
The spectral range used for analysis was selected to be $\lambda > 0.53$\,$\mu$m, because of the
transmittance curve of the GG495 order-blocking filter and $\lambda < 0.88$\,$\mu$m, because of strong
fringing above this wavelength. 
The resulting spectra are shown in Fig.~\ref{refl}. Regions with strong H$_2$O and O$_2$ telluric
absorption are indicated with $\vdash\hspace{-1pt}\dashv$ symbols in the bottom of the plot shown in
the left panel, showing the reflection spectrum obtained at WHT. We included these wavelength
intervals in the analysisbecause
the general
trend of the spectrum is not dramatically affected by these regions. 
In spite of an aggressive binning, measurement uncertainties remain considerable in the FoReRo2 data,
especially at wavelengths longer than 0.7 $\mu$m. However, we note that both spectra show a strong
negative slope shortward of 0.65\,$\mu$m where the S/N is highest; this bolsters
confidence that the WHT spectrum, in particular, is reliable and does not suffer from significant
systematic effects.

\subsection{Photometry -- FoReRo2 and T1m} 
All imaging data had their instrument signatures removed
the same way as the spectral data. Standard \textsc{daophot} aperture photometry with aperture size of 
2$\times$FWHM was performed. Standard stars in the field of view from the APASS
\footnote{https://www.aavso.org/apass} catalogue were used for absolute calibration.

\section{Results}
\label{results}
\subsection{Taxonomy}
\begin{figure*}
\includegraphics[width=\columnwidth]{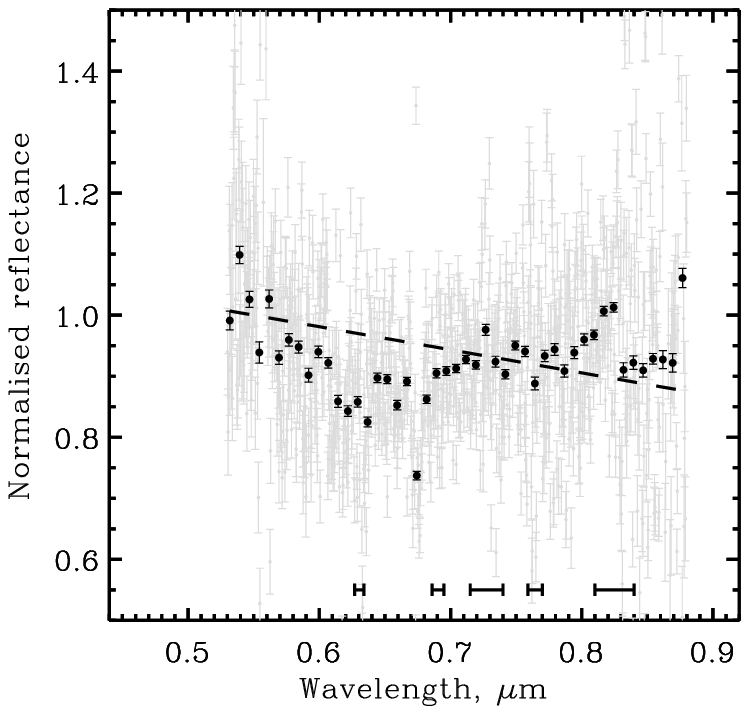} \includegraphics[width=\columnwidth]{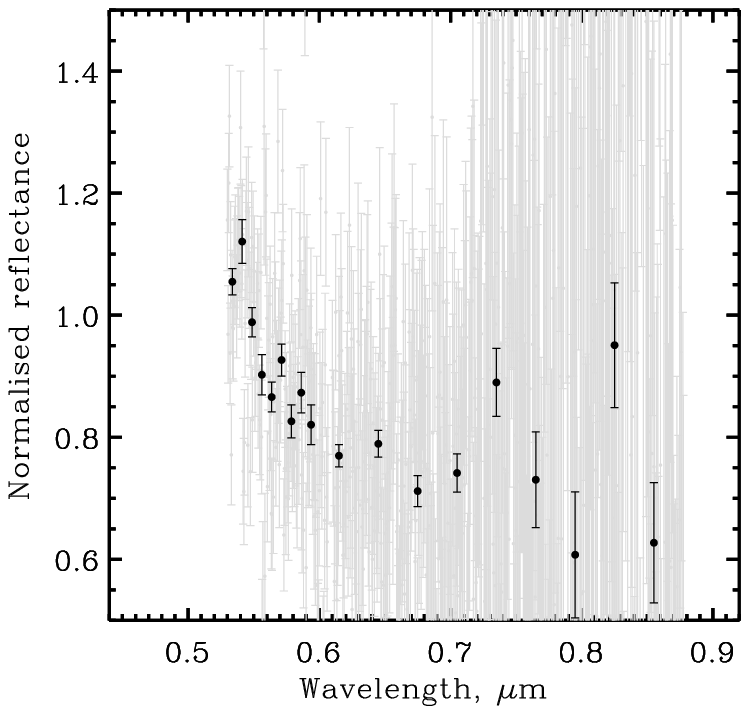}
\caption{Reflectance spectra of the asteroid (121514) 1999 UJ7. {\it Left panel}:  Reduced spectrum
  (grey) and the same data rebinned in 7.5\,nm steps (black) obtained with ACAM at WHT. Regions with
  strong H$_2$O and O$_2$ telluric absorption are denoted with $\vdash\hspace{-1pt}\dashv$.
  {\it Right panel}:  Spectrum obtained with FoReRo2@2mRCC. See text for details.}
\label{refl}
\end{figure*}

We determined the taxonomy of UJ7 by applying (i) slope-fitting and (ii) $\chi^{2}$ goodness-of-fit
ranking based only on the ACAM spectrum (left panel of Fig.~\ref{refl}) because the Rozhen spectrum
is of a much poorer quality. 

First, we estimated the slope $\gamma$ of the spectrum by fitting a function of the form
$r=1.0+\gamma(\lambda-0.55)$ to the data, where $\lambda$ is the wavelength in $\mu$m
\citep{BusBinzel2002}.
The best-fit slope was found to be $\gamma=-0.378^{+0.024}_{-0.025}$\,${\mu\mbox{m}}^{-1}$; this value is consistent
with, although at the high end of, a B-class taxonomy, which is a subgroup of the big, low-albedo C-complex
\citep{BusBinzel2002}. These asteroids are generally thought to be primitive, volatile-rich or
aqueously altered remnants from the early solar system. Although we computed the slope based on
a slightly narrower ($\gtrsim$70\%) wavelength range than in \citeauthor{BusBinzel2002}, 
the resulting value is not affected by a large uncertainty. As noted above,
the NEOWISE survey \citep{Nugent.et.al2015} obtained a low geometric albedo value of
0.053$\pm$0.034 and a diameter of 2.29$\pm$0.49\,km for this asteroid. In spite of its relatively
large albedo uncertainty, the nominal albedo value is consistent with a C-complex
classification.

The next step was to compute curve matching of the obtained spectrum using all known taxonomic classes
from \citet{BusBinzel2002}. 
Table~\ref{taxrank} shows the top five matches using the cost functions \citep{M4AST}

\begin{equation}
\begin{aligned}
\chi^2_{\mu}=\sum\limits_i^N{e_{i}^2 \over \mu_i},
\end{aligned}
\hspace{1cm}
\begin{aligned}
\Phi_{\rm std}={1 \over N}\sqrt{\sum\limits_{i}^{N}{(e_i-\bar{e})^2}},
\end{aligned}
\label{phi}
\end{equation}
where $e_i=x_i-\mu_i$ and $\bar{e}$ its mean value, $x_i$ is the measurement at the $i^{\rm th}$
wavelength position, and $\mu_i$ is the model evaluation at the same location. 

\begin{table}
\centering
\caption[Taxonomic ranking of the UJ7 spectrum]{Taxonomic ranking of the UJ7 spectrum.}
\label{taxrank}
\begin{tabular}{lclc}
\hline 
\hline      
\multicolumn{1}{c}{Taxonomic} &  & \multicolumn{1}{c}{Taxonomic} & \\
\multicolumn{1}{c}{class}        & $\chi^2_{\mu}$ & \multicolumn{1}{c}{class}   & $\Phi_{\rm std}$\\
\hline
\hspace{15pt}Ch &  0.326 & \hspace{15pt}Ch    & 0.0064\\
\hspace{15pt}B      & 0.377 & \hspace{15pt}Cb    &  0.0068\\
\hspace{15pt}Cgh & 0.459 & \hspace{15pt}Cgh  &  0.0072\\
\hspace{15pt}Cb    & 0.478 & \hspace{15pt}C      &  0.0075\\
\hspace{15pt}C      & 0.496 & \hspace{15pt}B      &  0.0083\\
\hline
\end{tabular}
\end{table}

Both cost functions assign the highest rankings to the C-complex taxonomies, including the C-class
itself and its four distinct subclasses Ch, Cb, B, and Cgh. The Ch class was found to give the best
match of our data.
The ranking order is different for the remaining four places. In particular, the $\chi^2_{\mu}$ assigns
the second place to the B taxonomy, while $\Phi_{\rm std}$ assigns B to the fifth place.

\begin{figure*}
\includegraphics[width=\columnwidth]{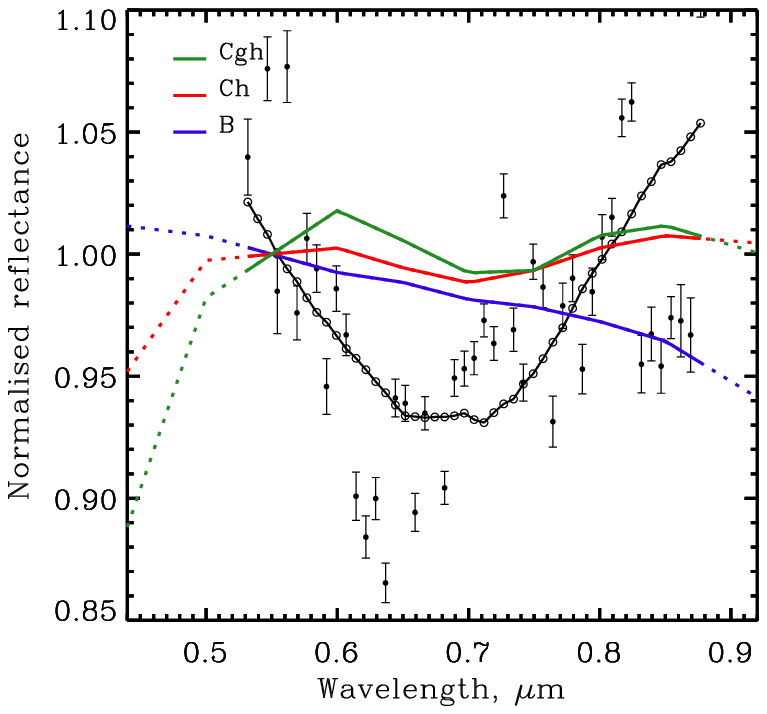} \includegraphics[width=\columnwidth]{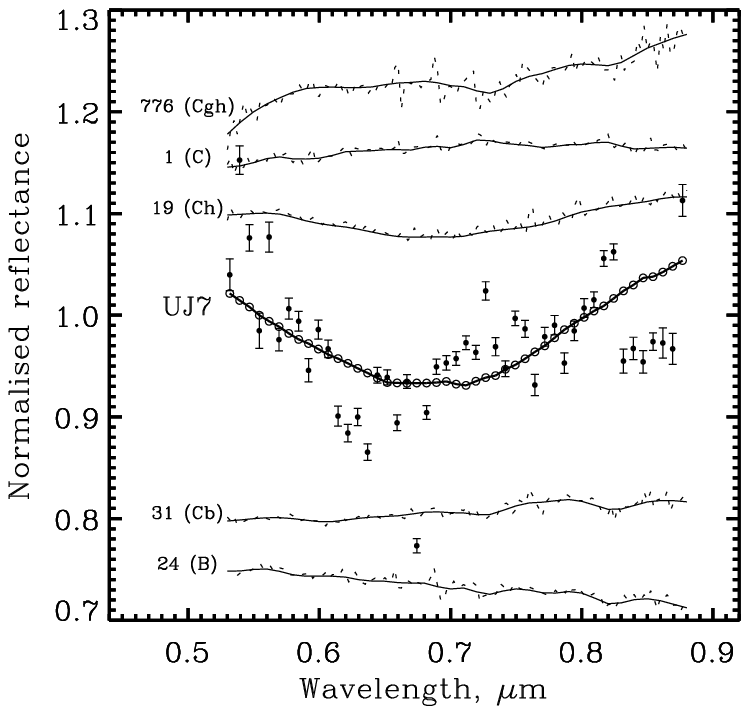}
\caption{{\it Left panel}: Average spectra of Ch, Cgh, and B classes from \citet{BusBinzel2002}
  (red, green, and blue lines, respectively) together with the binned spectrum of UJ7 (black dots). 
 The solid colour lines show the observed wavelength coverage of the average database spectra and 
  the solid line with open circles represents least-squares (Savitzky-Golay) polynomial 
  smoothing filter with window width equal to 47 of the observations. 
  {\it Right panel}: Comparison of the UJ7 spectrum (the same binned and smoothed as on the left panel) 
  with spectra of other asteroids (thin lines and dots) belonging to the taxonomic classes in Table~\ref{taxrank}.
  The spectra of the various asteroids represented in this figure are all normalised to 0.55\,$\mu$m, but each spectrum 
  (apart from that of UJ7) has been shifted up or down by some amount for the sake of clarity.}
\label{Tax}
\end{figure*}

A comparison of the UJ7 spectrum with average spectra for the Ch, B, and Cgh classes
is shown in the left panel of Fig.~\ref{Tax}.
The spectra of representative asteroids of the same classes as the top five matches 
presented in Table~\ref{taxrank} are compared with UJ7 spectrum as well (see the left panel of Fig.~\ref{Tax}).
UJ7 exhibits a strong absorption feature centred shortward of 0.7\,$\mu$m. This bears some
resemblance with a feature seen in Ch and Cgh spectra, but in the case of UJ7 we have a much wider
and deeper absorption feature and a maximum depth is located at a somewhat shorter wavelength.
The exact morphology (centre, depth, and width) of this spectral feature tends to vary a little
according to a more or less aggressive smoothing of the data.
We find that the centre of the absorption for different smoothing options is 
between 0.65 and 0.68\,$\mu$m, so it is only weakly dependent on data smoothing.
In any case, we find that this spectral feature is too wide and 
sharp to be a spurious telluric line artefact (see  Fig.~\ref{refl}).

The presence of a 0.7\,$\mu$m feature on asteroid spectra has been considered by many authors as
diagnostic of the presence of hydrated minerals on the surface
\citep{Lebofsky1980,VilasGaffey1989,Vilas.et.al1994}.
Among the asteroids plotted as a comparison in the right panel of Fig.~\ref{Tax}, 
(19) Fortuna (Ch-class) bears a better resemblance to our spectrum of UJ7. The 
absorption feature seen in Fortuna's spectrum, however, is significantly shallower than
in the case of UJ7, although it seems to be centred at a similar wavelength, shortward of 0.7 $\mu$m.
A much narrower absorption feature, centred around 0.7\,$\mu$m, can be seen also
in the spectrum of the Cgh-class asteroid (776) Berbericia, but the drop-off shortward of 0.6\,$\mu$m
is not present in this case, nor in the case of the spectrum of (1) Ceres, which is also shown in
the figure. To summarise, the reflectance spectrum of UJ7 bears some resemblance with 
the Ch subclass of the C-complex, but in some respects UJ7 might
represent a new and previously unknown case that can possibly be interpreted as the prototype
of a new subclass. 

The smoothed data on Fig.~\ref{Tax} is used just to show the global trend of the UJ7 spectrum 
and to find a better normalisation factor for the noisy data, but not for taxonomic ranking. 
The smoothing was carried out using least-squares (Savitzky-Golay) polynomial smoothing filter applied to the 
binned spectrum using the \textsc{idl} routine \textsc{poly\_smooth} from NASA IDL Astronomy 
User's Library\footnote{https://idlastro.gsfc.nasa.gov}. 

Of course, we cannot firmly state that the absorption feature observed in UJ7 must be forcedly
considered as diagnostic of water of hydration. We note, however, that the 0.7\,$\mu$m feature
detected in asteroid spectra varies considerably in different cases, typically between 0.67\,$\mu$m
and 0.72\,$\mu$m for the location of its centre, and between 
1\% and 4\% in depth. In some individual cases it can reach locations down to 0.65\,$\mu$m and
depths up to 6\% \citep[see][Figure 13]{Fornasier.et.al2014}. A tentative interpretation of our data
may therefore be that the observed feature is diagnostic of the presence of hydrated
minerals, and its unusual attributes may be only partly due to incomplete telluric line removal
combined with the coarse resolution of the spectrum.

The peculiar properties of the strong absorption feature found in the UJ7 spectrum
  can be thoroughly real and represent a first example of a previously unknown subclass
  of the C-complex. This might be related to the extended residence time of this asteroid at 1.5\,au,
  namely closer to the Sun than any classical main belt asteroid.

Laboratory experiments in which meteorite powders were subjected to varying
degrees of heating show that this feature is destroyed at temperatures
$>$400\,$^\circ$C \citep{Hiroi.et.al1996}. Assuming energy balance, a
typical subsolar point temperature for UJ7 is $T$ = $[(1 -  A)*S/\eta
\epsilon \sigma ]^{1/4}$ \citep{LebofskySpencer1989}, where $A$ is the 
Bond albedo; $S$ the solar energy influx
$S=S_{0}/a^{2}$, where $S_{0}$ is the solar constant; $a$ the semi-major
axis of the asteroid; $\eta$ the beaming parameter; $\sigma$ Boltzmann's
constant; and $\epsilon$ the infrared emissivity. 
\citet{Nugent.et.al2015} obtained $\eta=0.95 \pm 0.18$ for this
asteroid by applying the Near Earth Asteroid Thermal Model
\cite[NEATM;][]{Harris1998} to NEOWISE survey data. Although we are aware of
alternative, potentially more accurate, methods to constrain asteroid
surface properties from thermal observations
\cite[e.g.][]{Lebofsky1986,Spencer1990} we adopted the NEATM
values for $p_{V}$ and $\eta$ as they are obtained self-consistently from the model. 
Likewise, we adopt $\epsilon=0.9$ to maintain consistency with the NEATM. 

For $a=1.52$\,au, $A=0.393 p_{V}$ \cite[Eqs~33 \& A7 of][where we have used G=0.15]{Bowell.et.al1989} and $\eta$ equal to the \citeauthor{Nugent.et.al2015} value, we obtain $T$=331\,K. 
An alternative thermal solution that uses WISE W3 and W4 band data \citep{Mainzer.et.al2012} in addition to the W1 and W2 band data used in \citeauthor{Nugent.et.al2015} is reported in \citet{AliLagoaDelbo2017}, giving $p_{V}=0.041$ and $\eta=1.08$.
Carrying out the same calculation as above, we obtain a slightly lower temperature, 321\,K. In any case, this temperature varies depending on the eccentricity
history of the orbit; for $e\lesssim 0.1$ expected for stable Martian
Trojans the variation may be as high as 302$-$368\,K if we also consider
the reported 1$\sigma$ uncertainty for $\eta$ in \citeauthor{Nugent.et.al2015}. Therefore, the presence of a water-of-hydration feature on UJ7 is consistent with 
this primitive body having been in its present orbit or
further from the Sun for $\sim$4\,Gyr.

For completeness, we also note that a feature at $0.654$\,$\mu$m has been identified in spectra 
of Phobos and Deimos by \citet{Fraeman.et.al2014}.
Those authors obtained a best-fit model of the feature for a mixture of iron particles of various
sizes embedded in a neutral silicate matrix. That feature was superimposed on a red-sloped continuum that is
not apparent in our data, therefore we consider this a less likely interpretation for the UJ7 spectrum.
The authors found a few asteroids in the Vilas asteroid spectral catalogue \citep{Vilas} that have 
the same absorption feature that is also superimposed on the red-sloped continuum. They ascribe this 
feature to Fe-bearing phyllosilicates that are almost always accompanied by additional absorption at 
3\,$\mu$m associated with hydration or hydroxylation. Interestingly all these asteroids are further away 
than 3\,au, which may point us to a possible origin of UJ7.

\subsection{Spin state}
To determine the spin rate of the asteroid, the time series of the absolute calibrated magnitude of
UJ7 was fitted to a harmonic function. We used a single-period Fourier series of the form
\begin{equation}
F^m(t) = C_{00}+\sum\limits^m_{j=1}\left[ C_{j0}\cos{2\pi j \over P_{\psi}}t + S_{j0}\sin{2\pi j \over P_{\psi}}t \right],
\end{equation}
where $m$ is the series order (we used third order), $C_{00}$ is the mean reduced magnitude, $C_{j0}$ and $S_{j0}$ are the 
Fourier coefficient,  $P_{\psi}^{-1}$  represents the frequency, and $t$ is the time.

\begin{figure*}
\includegraphics[width=0.67\columnwidth]{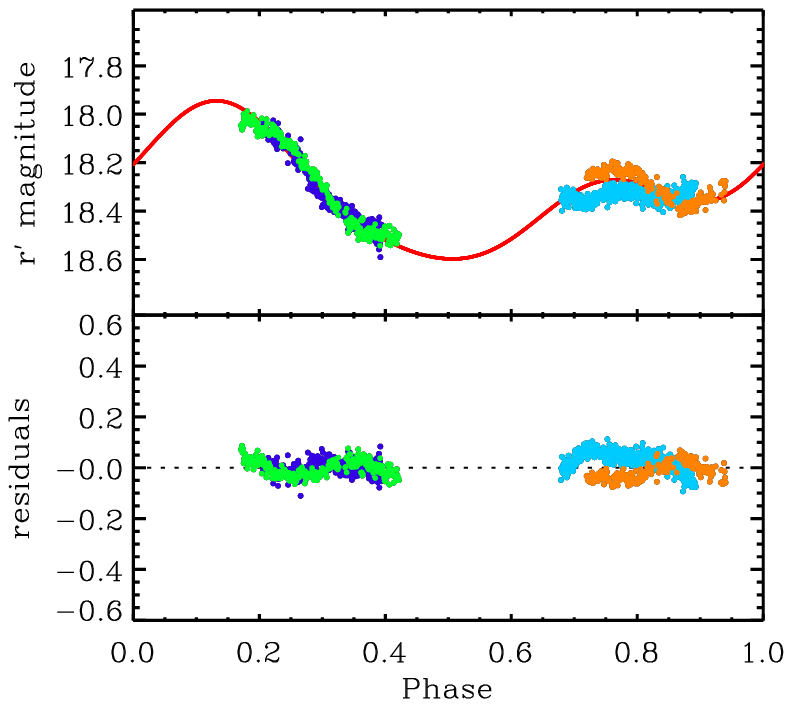} \includegraphics[width=0.67\columnwidth]{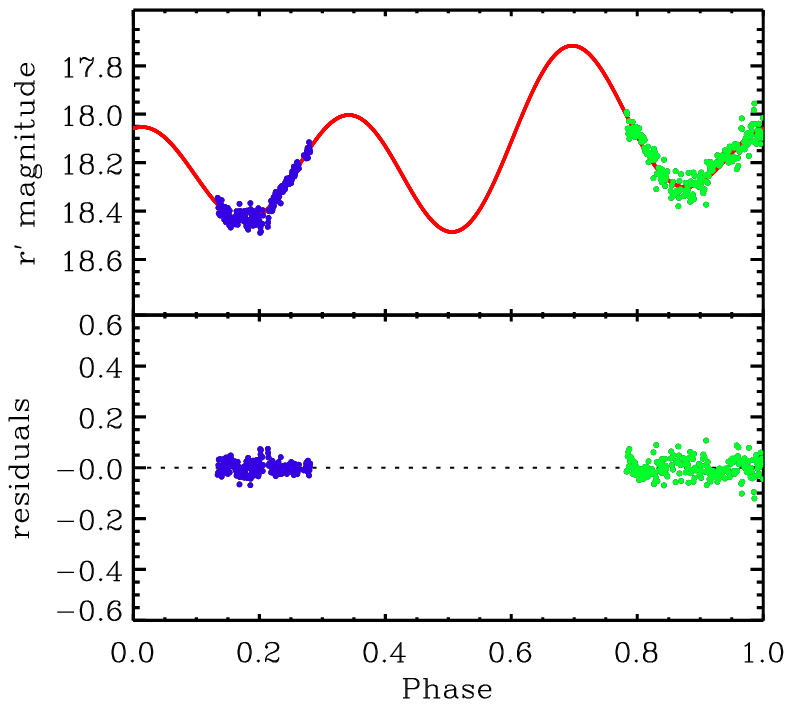} \includegraphics[width=0.67\columnwidth]{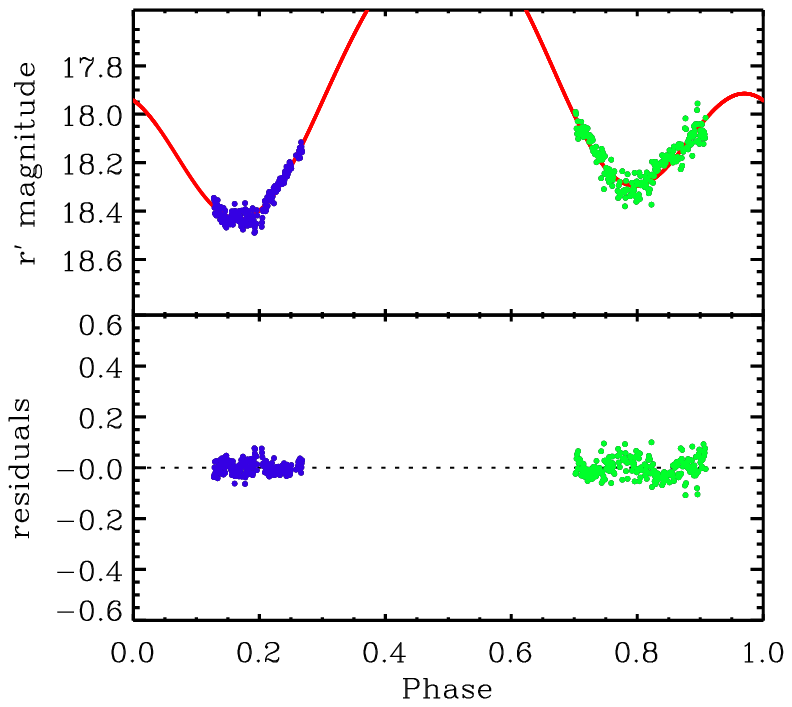}
\caption{Fourier fits to the photometric data of the asteroid (121514) 1999 UJ7. Different colours correspond to
  data from different nights. {\it Left panel}: Light curve with fitted 3rd order Fourier polynomial in red
  (top panel) and the residuals (bottom panel) using data from 1, 2, 5, and 6 Dec 2016. The obtained period is
  1.936$\pm$0.002\,d. 
  {\it Middle panel}: As for the left panel but fitting data from 29 Dec 2016 and 01 Jan 2017. The obtained period
  in this case is 1.848$\pm$0.028\,d. {\it Right panel}: As in middle panel but keeping the period fixed at
  $P=$1.936\,d.}
\label{Dec}
\end{figure*}

We did not succeed to obtain a satisfactory fit to the entire dataset, so we divided the data into two groups:
one for early and the other for late December and fit these separately. This yielded periods of 1.936\,d for the
early December and 1.848\,d for the late December data (Fig.~\ref{Dec}). We note that the model light curves appear
different for the two fits.

\begin{figure*}
\includegraphics[height=0.83\columnwidth]{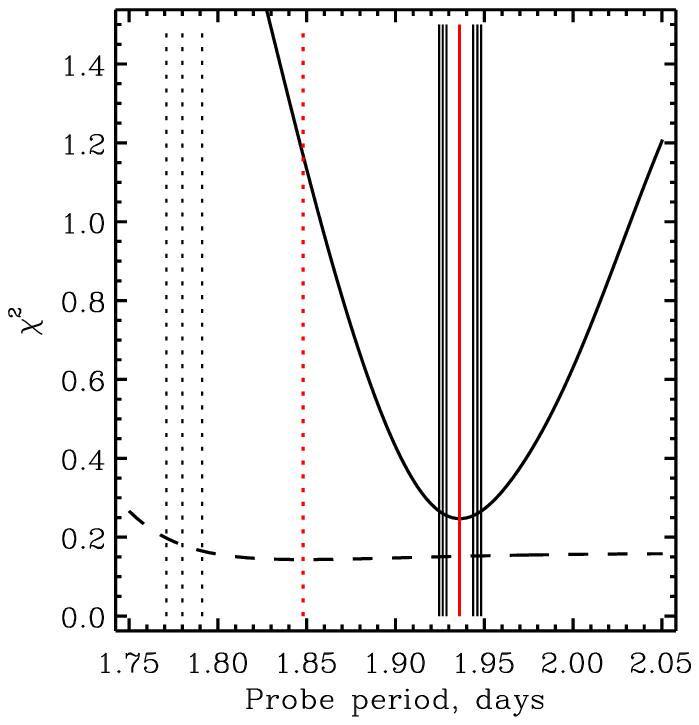} \hspace*{1.5cm} \includegraphics[width=\columnwidth]{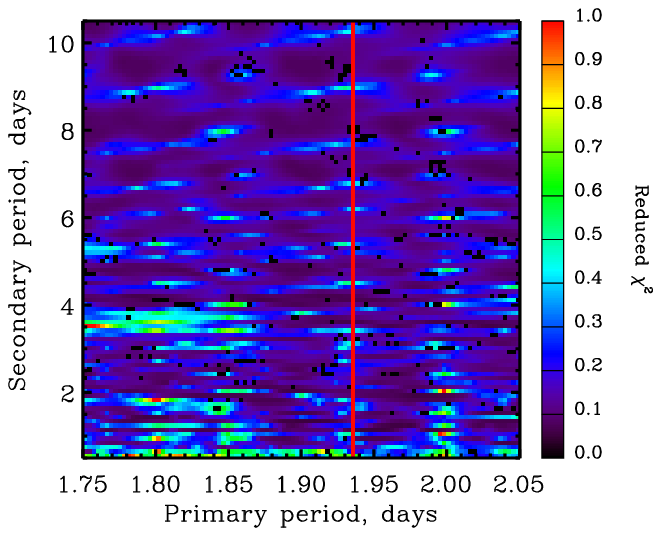}
\caption{{\it Left panel}: $\chi^{2}$ of a Fourier polynomial fit varying the period between 1.75 and 2.05\,d. The solid curve represents results from fitting data from 1, 2, 5, and 6 Dec 2016, while the dashed curve uses data from 29 Dec 2016 and 01 Jan 2017. The red solid and dotted vertical straight lines indicate best-fitted periods of 1.936$\pm$0.002 and 1.848$\pm$0.028\,d, respectively. The black solid and dotted lines represent 1, 2 and 3$\sigma$ uncertainties for each fit as follows: $\pm7.5$, 
$^{+9.6}_{-9.9}$ and $\pm12$ $\times 10^{-3}$\,d. {\it Right panel}: $\chi^{2}$ distribution for a two-period Fourier polynomial fit. The single-period solution of 1.936\,d is indicated by the vertical red line.}
\label{chisq}
\end{figure*}
\begin{figure}
\hspace*{-0.7cm}
\includegraphics[width=1.05\columnwidth]{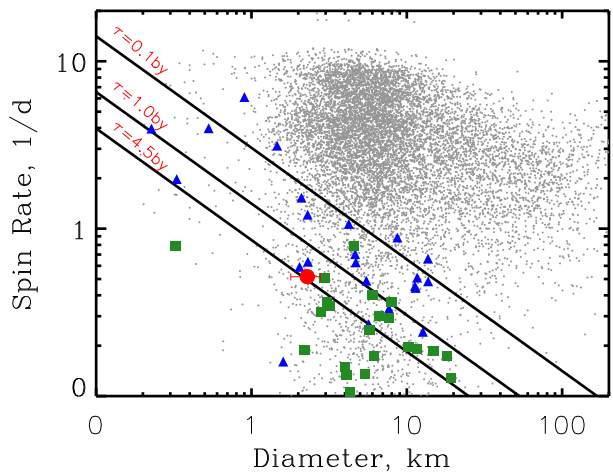}
\caption{Asteroid spin rate vs.~diameter. The datum for UJ7 is presented as a red dot. All NPA and PA asteroids with known spin rate and size are presented with green squares and blue triangles, respectively. The lines represent damping timescales of 4.5, 1, and 0.1 (bottom to top) for NPA rotation.}
\label{diaper}
\end{figure}

Plotting the $\chi^2$ statistic (Fig.~\ref{chisq}), we find that the fit for the late December data is somewhat
shallow and broad compared to that for early December; the latter solution lies within 1$\sigma$ of the
former. This suggests that the period is constrained mainly from the early December data. We tested this by
again fitting the late December data but using a fixed value $P=1.936$\,d for the period while allowing the
remaining parameters to vary in the fit (right panel of Fig.~\ref{Dec} and see caption). Although the fit itself
is satisfactory, the best-fit model appears, not unexpectedly, to be significantly different, in terms of
light-curve morphology, with respect to what was previously. A simple
explanation might be that the late December dataset is, by itself, insufficient to  constrain the
light-curve parameters. On the other hand, if this was the case, and the period derived by the only early December
data was correct, we could not explain the difficulty encountered in fitting all the data together. It seems
rather that there is some significant change of morphology of the light curve taking place in only a few
weeks. A possible reason could be the interplay of an asymmetrical shape and a change
of aspect angle of the asteroid (the angle between the rotational axis and line of the sight).
It was shown long ago \citep{Cellinoetal89} that significant changes in light-curve morphology can be
due to shape effects in a large variety of cases. 
On the other hand, the time span between the first and the last observations seems too short to
justify big light-curve morphology changes. 
Another possibility is that a change in the rotational state of the asteroid may
have occurred over an interval of $\sim1$ month, implying that the asteroid is not in principal-axis (PA) rotation.
On Figure~\ref{diaper} we show spin rate versus~diameter for more than 10\,000 asteroids
\footnote{JPL Solar System Dynamics Database: https://ssd.jpl.nasa.gov/ sbdb\_query.cgi}.
All non-principal-axis (NPA) and PA rotation asteroids with known spin rate and size are also 
presented with green squares and blue triangles, respectively. 
The straight  lines illustrate damping timescales of 4.5, 1, and 0.1\, (bottom to top) for 
NPA rotation using the relation $P\approx  C D^{2/3}\tau^{1/3}$ from \citet{Harris1994}, 
where $C\sim17$ (uncertain by a factor of $\sim$2.5), $P$ is the rotation period in hr, $D$ the diameter 
in km, and $\tau$ the damping timescale in billions of years.
Plotting the position of the asteroid on this diagram shows that UJ7 is found  near the 
NPA rotators and near some of PA rotators, but also 
near the 4.5\,by damping timescale line. 
The possibility that this asteroid can be an NPA rotator cannot
be ruled out for an object of its size. This would explain the difficulty in obtaining a good single-period
fit for the entire one-month data arc. On the other hand, we have to stress that the location of UJ7 in the
spin rate - size plot cannot be a strong argument, suggesting only a partial indication about it. 
The NPA rotation implies the presence of a secondary period in the light curve. Therefore we
searched for a two-period solution by fitting a Fourier series as in \citet{Pravec.et.al2005} while scanning
through solution space for primary and secondary periods in the domain $[1.75,2.05]\times[0.5,10.5]$\,d. The
distribution of $\chi^{2}$ for this search is shown in the right panel of Fig~\ref{chisq}. Several local minima
are apparent but no single solution stands out. In conclusion, we are unable to confirm that this asteroid is
in an NPA rotational state; a definite answer to this question will require better observational coverage of
its light curve. 

\section{Discussion}
\label{blabla}
\citet{Rivkin.et.al2003} classified the asteroid as X-class based on a featureless
spectrum of positive visible slope (their Fig.~6). Our spectrum, however, shows a clearly
negative slope at the shortest wavelengths.
Our results confirm the \citeauthor{Rivkin.et.al2003} conclusion that the mineralogy of UJ7
is unlike that of its neighbours at $\mbox{L}_{5}$. However, our new spectrum supports a 
spectral classification that is consistent with the NEOWISE visible albedo, suggesting that this asteroid
has suffered minimal to no thermal, or geological, processing since the epoch of its
formation.

The difference between our spectrum and that obtained in the past by \citet{Rivkin.et.al2003}
might be possibly interpreted as evidence for rotational variability across the surface of the asteroid.
Visible slope variability has also been documented for C-complex near-Earth objects: (153591) 2001
$\mbox{SN}_{263}$ \cite[D=1\,km][]{Perna.et.al2013}, where the slope varies from flat to strongly
negative (their Fig.~2) and (175706) 1996 $\mbox{FG}_{3}$ \citep[D=1.7\,km]{Perna.et.al2014}, where
the slope varies from positive to weakly negative (their Fig.~3).
In addition, a set of Main Belt asteroids investigated by \citet{Fornasier.et.al2014} also shows
some significant slope variability in the visible region (their Table 5), even for different spectra
acquired from the same asteroid, as in the case of (1467) Mashona, showing both a positive and
negative slope. Therefore, a strong rotational dependence of the spectral slope for UJ7, while
unusual, is not unprecedented.

\citet{Binzel.et.al2015} interpreted variability in the observed spectral slope of asteroid (109155)
Bennu, target of the OSIRIS-REx mission, in terms of regolith grain size variations
across its surface. According to their model, the observed spectral slope would depend on the
existence of an equatorial ridge created by a YORP spin-up. In the case of UJ7, however, the slow
rotation argues against the existence of such a ridge.

The spectral slope of primitive asteroid surfaces may also be modified
by space weathering \citep{Moroz.et.al1996,Lantz.et.al2013,Matsuoka.et.al2015}. In this context, the
existence of surface units of different colours would imply a resurfacing mechanism at work, perhaps
a cratering impact that created a patch of fresh material on the surface of UJ7.

The difference between our spectrum of UJ7 and the spectrum obtained in
the past by \citet{Rivkin.et.al2003} might be due to some spurious effect, which is unlikely 
because the spectral feature is very similar observed with both the ACAM and FoReRo2 instruments. 
In this case, however, we
do not deal with a simple difference in the spectra slope, but we have a different overall
morphology of the reflectance spectrum.

Ch- and Cgh-type asteroids may typically display a sharp spectral feature at $\sim$3\,$\mu$m
\citep{TakirEmery2012}.
These asteroids are considered to be likely parent bodies of the most common class of hydrated meteorites,
namely CM chondrites \cite[][and references therein]{Vernazza.et.al2016}. Unlike other CC meteorites,
the CM parent bodies seem to have experienced only mild heating ($<$ 400\,$^\circ$C) during their lifetime,
to preserve the $0.7$\,$\mu$m feature. The $\sim$3\,$\mu$m feature survives heating up to 700\,$^\circ$C
\citep{Hiroi.et.al1996}, therefore in the case in which the absorption band in the spectrum of UJ7 is due to the presence
of hydration, we can predict that this asteroid should also exhibit a  $\sim$3\,$\mu$m absorption feature; this hypothesis can be verified or contradicted by future observations in the IR.

At the same time, we consider it unlikely that this asteroid has retained any traces of its original cache
of volatiles in its interior. The characteristic timescale to achieve a uniform temperature throughout the
asteroid is $t_{h}=\frac{1}{12} {\left(D \rho C / \Gamma\right)}^{2}$, where $D$ is the diameter, $\Gamma$ the
thermal inertia, $\rho$ the bulk density, and $C$ the specific heat capacity \citep{Busarev.et.al2017}.
For $D=2300$\,m, $\Gamma=200$\,J $\mbox{K}^{-1}$\,$\mbox{m}^{-2}$\,$\mbox{s}^{-1/2}$, $\rho=1300$\,kg $\mbox{m}^{-3}$,
and $C=600$\,J\,$\mbox{kg}^{-1}$\,$\mbox{K}^{-1}$, we obtain $t_{h}\simeq 2 \times 10^{5}$\,yr, which is four orders of
magnitude shorter than the likely residence time of the asteroid at 1.5\,au.
 
According to the so-called Grand-Tack \citep{Walsh.et.al2011} model of the early solar system evolution,
volatile-rich planetesimals originally accreted between and
beyond the original orbits of the giant planets were scattered inwards and became embedded into the asteroid
belt as a consequence of a complicated path of migration of Jupiter. Water ice that has been detected on the surfaces of outer
belt asteroids \citep{Campins.et.al2010,RivkinEmery2010,Licandro.et.al2011} and objects of similar composition
may have been responsible for the delivery of water to the Earth
\citep{Morbidelli.et.al2000,Morbidelli.et.al2012,Izidoro.et.al2013,RaymondIzidoro2017}. On the other hand,
\citet{Polishook.et.al2017}, in arguing for a Martian origin for most of the L$_5$ Mars Trojans, showed that
Mars Trojans were efficiently locked into place during the final stages of terrestrial planet formation,
when encounters with planetary embryos may have caused chaotic wandering of the Martian orbit. Therefore,
the existence of a member of the C-complex located in L$_4$, as concluded from the spectral slope and
taxonomic matching, constitutes direct evidence for the presence of geologically unprocessed objects not
originating in the region of accretion of Mars, as opposed to the case of the Eureka family
\citep{Polishook.et.al2017} in L$_5$. In this context, a possible hypothesis is that UJ7 may have been originated
in a region near or beyond the snow line. During the early chaotic phases predicted by the Grand Tack model, it
can have been perturbed in such a way as to finally achieve a low-eccentricity orbit at 1.5\,au, where it was
eventually captured as a permanent Mars Trojan when the planetary embryos in Mars-crossing orbits
started to disappear. In particular, it is known that after the phase described by the Grand-Tack model,
many small objects present in the outer regions of the solar system continued to be subject to strong
perturbations, until the end of the so-called Late Heavy Bombardment era. This epoch of transition between the
early planetary migrations and the final establishment of the outer planets in their current orbital
locations is described by the so-called Nice model \citep{Nice}. 
An important implication of the scenario described above is that the original influx of C-complex asteroids moving 
into Mars-like orbits must have been not only significant, if we consider that the estimated mass of UJ7 
accounts for $\sim$50\% of the current total mass budget in the Trojan clouds, but also took place in very 
early epochs when the capture of objects into Trojan orbits was easier. 

Regarding future observational prospects, the next two apparitions of UJ7 in 2018/19 and 2020/21 are slightly
more favourable than in 2016/17. The asteroid will be at  $V<18$ at its brightest. Observation modes such as
spectroscopy requiring long dwell times are facilitated during periods when the asteroid is near stationary.
For UJ7, this did not occur in 2016/17 but will occur in Dec 18/\,Jan 19 and April/May 21 respectively.
High S/N spectroscopy both in the visible and the near-IR would allow the search for additional evidence of
hydrated minerals and would better constrain the surface heterogeneity with respect to what was done in this work and
by \cite{Rivkin.et.al2003}. Furthermore, additional photometric coverage will be important to confirm a possible
complex rotational state of this asteroid. Finally, our understanding of UJ7 and other similar asteroids will
presumably benefit from the
ongoing {\it Hayabusa 2} and {\it OSIRIS-REx} missions, which will conduct  in situ investigations and
return samples from two C-complex asteroids: (162173) Ryugu and (101955) Bennu
\citep{Clark.et.al2011,Perna.et.al2017}.

\begin{acknowledgements}
This work was supported via a grant (ST/M000834/1) from the UK Science and Technology Facilities Council.
The 4.2 m William Herschel Telescope and its service programme SW2016b13 are operated on the island of 
La Palma by the Isaac Newton Group of Telescopes in the Spanish Observatorio del Roque de los Muchachos 
of the Instituto de Astrof\'{i}sica de Canarias. 
We gratefully acknowledge observing grant support from the Institute of Astronomy and Rozhen National Astronomical 
Observatory, Bulgarian Academy of Sciences. 
This research was made possible through the use of the AAVSO Photometric All-Sky Survey (APASS), 
funded by the Robert Martin Ayers Sciences Fund. 
This publication also makes use of data products from NEOWISE, which is a project of the JPL/California Institute of Technology, funded by the Planetary Science Division of NASA. 
\end{acknowledgements}

%
%

\bibliographystyle{aa} 
\bibliography{MarTr} 

\begin{thebibliography}{55}
\expandafter\ifx\csname natexlab\endcsname\relax\def\natexlab#1{#1}\fi

\bibitem[{{Al\'{i}-Lagoa} \& {Delbo}(2017)}]{AliLagoaDelbo2017}
{Al\'{i}-Lagoa}, V. \& {Delbo}, M. 2017, A\&A, 603, id.A55

\bibitem[{{Binzel} {et~al.}(2015){Binzel}, , {DeMeo}, \& {15
  co-authors}}]{Binzel.et.al2015}
{Binzel}, R.~P., , {DeMeo}, F.~E., \& {15 co-authors}. 2015, Icarus, 256, 22

\bibitem[{{Borisov} {et~al.}(2017){Borisov}, {Christou}, {Bagnulo}, {Cellino},
  {Kwiatkowski}, \& {Dell'Oro}}]{BorisovMarTr2017}
{Borisov}, G., {Christou}, A., {Bagnulo}, S., {et~al.} 2017, MNRAS, 466, 489

\bibitem[{{Bowell} {et~al.}(1989){Bowell}, {Hapke}, {Domingue}, {Lumme},
  {Peltoniemi}, \& {Harris}}]{Bowell.et.al1989}
{Bowell}, E., {Hapke}, B., {Domingue}, D., {et~al.} 1989, in Asteroids II,
  524--556

\bibitem[{{Britt} {et~al.}(1989){Britt}, {Bell}, {Haack}, \&
  {Scott}}]{Bell.et.al1989}
{Britt}, D.~T., {Bell}, J.~F., {Haack}, H., \& {Scott}, E.~R.~D. 1989, In:
  Asteroids II, University of Arizona Press,, 27, 921

\bibitem[{{Britt} {et~al.}(1992){Britt}, {Bell}, {Haack}, \&
  {Scott}}]{Britt.et.al1992}
{Britt}, D.~T., {Bell}, J.~F., {Haack}, H., \& {Scott}, E.~R.~D. 1992,
  Meteoritics, 27, 207

\bibitem[{{Bus} \& {Binzel}(2002)}]{BusBinzel2002}
{Bus}, S.~J. \& {Binzel}, R.~P. 2002, Icarus, 158, 146

\bibitem[{{Busarev} {et~al.}(2017){Busarev}, {Makalkin}, {Vilas}, {Barabanov},
  \& {Scherbina}}]{Busarev.et.al2017}
{Busarev}, V.~V., {Makalkin}, A.~B., {Vilas}, F., {Barabanov}, S.~I., \&
  {Scherbina}, M.~P. 2017, Icarus, in press

\bibitem[{{Campins} {et~al.}(2010){Campins}, {Hargrove}, {Pinilla-Alonso},
  {Howell}, \& {et al.}}]{Campins.et.al2010}
{Campins}, H., {Hargrove}, K., {Pinilla-Alonso}, N., {Howell}, E.~S., \& {et
  al.} 2010, Nature, 464, 1320

\bibitem[{{Cellino} {et~al.}(1989){Cellino}, {Zappala}, \&
  {Farinella}}]{Cellinoetal89}
{Cellino}, A., {Zappala}, V., \& {Farinella}, P. 1989, Icarus, 78, 298

\bibitem[{{Christou}(2013)}]{Christou2013}
{Christou}, A.~A. 2013, Icarus, 224, 144

\bibitem[{{Clark} {et~al.}(2011){Clark}, {Binzel}, \& {13
  co-authors}}]{Clark.et.al2011}
{Clark}, B.~E., {Binzel}, R.~P., \& {13 co-authors}. 2011, Icarus, 216, 462

\bibitem[{{de la Fuente Marcos} \& {de la Fuente
  Marcos}(2013)}]{deLaFuenteMarcoses2013}
{de la Fuente Marcos}, C. \& {de la Fuente Marcos}, R. 2013, MNRAS, 432, 31

\bibitem[{{Desch}(2007)}]{Nice}
{Desch}, S.~J. 2007, ApJ, 671, 878

\bibitem[{{Emery} {et~al.}(2011){Emery}, {Burr}, \&
  {Cruikshank}}]{Emery.et.al2011}
{Emery}, J.~P., {Burr}, D.~M., \& {Cruikshank}, D.~P. 2011, Astronomical
  Journal, 141, id.~25

\bibitem[{{Fornasier} {et~al.}(2014){Fornasier}, {Lantz}, {Barucci}, \&
  {Lazzarin}}]{Fornasier.et.al2014}
{Fornasier}, S., {Lantz}, C., {Barucci}, M.~A., \& {Lazzarin}, M. 2014, Icarus,
  233, 163

\bibitem[{{Fraeman} {et~al.}(2014){Fraeman}, {Murchie}, {Arvidson}, {Clark},
  {Morris}, {Rivkin}, \& {Vilas}}]{Fraeman.et.al2014}
{Fraeman}, A.~A., {Murchie}, S.~F., {Arvidson}, R.~E., {et~al.} 2014, Icarus,
  229, 196

\bibitem[{{Harris}(1994)}]{Harris1994}
{Harris}, A.~W. 1994, Icarus, 207, 209

\bibitem[{{Harris}(1998)}]{Harris1998}
{Harris}, A.~W. 1998, Icarus, 131, 291

\bibitem[{{Hiroi} {et~al.}(1996){Hiroi}, {Zolensky}, {Pieters}, \&
  {Lipschutz}}]{Hiroi.et.al1996}
{Hiroi}, T., {Zolensky}, M.~E., {Pieters}, C.~M., \& {Lipschutz}, M.~E. 1996,
  Met.~Planet.~Sci., 31, 321

\bibitem[{{Izidoro} {et~al.}(2013){Izidoro}, {de Souza Torres}, {Winter}, \&
  {HaghiGHipour}}]{Izidoro.et.al2013}
{Izidoro}, A., {de Souza Torres}, K., {Winter}, O.~C., \& {HaghiGHipour}, N.
  2013, Astrophysical Journal, 767, 54

\bibitem[{{Jockers} {et~al.}(2000){Jockers}, {Credner}, {Bonev}, {Kisele},
  {Korsun}, {Kulyk}, {Rosenbush}, {Andrienko}, {Karpov}, {Sergeev}, \&
  {Tarady}}]{FoReRo2}
{Jockers}, K., {Credner}, T., {Bonev}, T., {et~al.} 2000, Kinematika i Fizika
  Nebesnykh Tel Supplement, 3, 13

\bibitem[{{Lantz} {et~al.}(2013){Lantz}, {Clark}, {Barucci}, \&
  {Lauretta}}]{Lantz.et.al2013}
{Lantz}, C., {Clark}, B.~E., {Barucci}, M.~A., \& {Lauretta}, D.~S. 2013,
  Astron.~Astrophys., 554, A138

\bibitem[{{Lebofsky}(1980)}]{Lebofsky1980}
{Lebofsky}, L.~A. 1980, Astronomical Journal, 85, 573

\bibitem[{{Lebofsky} \& {Spencer}(1989)}]{LebofskySpencer1989}
{Lebofsky}, L.~A. \& {Spencer}, J.~R. 1989, In: Asteroids II (R.~P.~Binzel,
  T.~Gehrels, and M.~S.~Matthews, eds.), Arizona University Press, Tucson, 128

\bibitem[{{Lebofsky} {et~al.}(1986){Lebofsky}, {Sykes}, {Tedesco}, {Veeder},
  {Matson}, {Brown}, {Gradie}, {Feierberg}, \& {Rudy}}]{Lebofsky1986}
{Lebofsky}, L.~A., {Sykes}, M.~V., {Tedesco}, E.~F., {et~al.} 1986, Icarus, 68,
  239

\bibitem[{{Licandro} {et~al.}(2011){Licandro}, {Campins}, {Kelley}, {Hargrove},
  {Pinilla-Alonso}, {Cruikshank}, {Rivkin}, \& {Emery}}]{Licandro.et.al2011}
{Licandro}, J., {Campins}, H., {Kelley}, M., {et~al.} 2011, Astron.~Astrophys.,
  525, A34

\bibitem[{{Mainzer} {et~al.}(2012){Mainzer}, {Grav}, {Masiero}, {Bauer},
  {Cutri}, {McMillan}, {Nugent}, {Tholen}, {Walker}, \&
  {Wright}}]{Mainzer.et.al2012}
{Mainzer}, A., {Grav}, T., {Masiero}, J., {et~al.} 2012, Astrophys.~J.~Lett.,
  760, article id.~L12

\bibitem[{{Matsuoka} {et~al.}(2015){Matsuoka}, {Nakamura}, {Kimura}, {Hiroi},
  {Nakamura}, {Okumura}, \& {Sasaki}}]{Matsuoka.et.al2015}
{Matsuoka}, M., {Nakamura}, T., {Kimura}, Y., {et~al.} 2015, Icarus, 254, 135

\bibitem[{{Morbidelli} {et~al.}(2000){Morbidelli}, {Chambers}, {Lunine},
  {Petit}, {Robert}, {Valsecchi}, \& {Cyr}}]{Morbidelli.et.al2000}
{Morbidelli}, A., {Chambers}, J., {Lunine}, J.~I., {et~al.} 2000,
  Met.~Planet.~Sci., 35, 1309

\bibitem[{{Morbidelli} {et~al.}(2012){Morbidelli}, {Lunine}, {O'Brien},
  {Raymond}, \& {Walsh}}]{Morbidelli.et.al2012}
{Morbidelli}, A., {Lunine}, J.~I., {O'Brien}, D.~P., {Raymond}, S.~N., \&
  {Walsh}, K.~J. 2012, Ann.~Rev.~Earth.~Planet.~Sci., 40, 251

\bibitem[{{Moroz} {et~al.}(1996){Moroz}, {Fisenko}, {Semjonova}, {Pieters}, \&
  {Korotaeva}}]{Moroz.et.al1996}
{Moroz}, L.~V., {Fisenko}, A.~V., {Semjonova}, L.~F., {Pieters}, C.~M., \&
  {Korotaeva}, N.~N. 1996, Icarus, 122, 366

\bibitem[{{Nugent} {et~al.}(2015){Nugent}, {Mainzer}, {Masiero}, {Bauer},
  {Cutri}, {Grav}, {Kramer}, {Sonnett}, {Stevenson}, \&
  {Wright}}]{Nugent.et.al2015}
{Nugent}, C.~R., {Mainzer}, A., {Masiero}, J., {et~al.} 2015, Astrophys.~J.,
  814, article id.~117

\bibitem[{{Perna} {et~al.}(2014){Perna}, {Alvarez-Candal}, {Fornasier},
  {Ka\v{n}uchov\'{a}}, {Giuliatti Winter}, {Vieira Neto}, \&
  {Winter}}]{Perna.et.al2014}
{Perna}, D., {Alvarez-Candal}, A., {Fornasier}, S., {et~al.} 2014,
  Astron.~Astrophys., 568, L6

\bibitem[{{Perna} {et~al.}(2017){Perna}, {Barucci}, \& {10
  co-authors}}]{Perna.et.al2017}
{Perna}, D., {Barucci}, M.~A., \& {10 co-authors}. 2017, Astron.~Astrophys.,
  599, L1

\bibitem[{{Perna} {et~al.}(2013){Perna}, {Dotto}, {Barucci}, {Fornasier},
  {Ka\v{n}uchov\'{a}}, {Giuliatti Winter}, {Vieira Neto}, \&
  {Winter}}]{Perna.et.al2013}
{Perna}, D., {Dotto}, E., {Barucci}, M.~A., {et~al.} 2013, Astron.~Astrophys.,
  555, id.A62

\bibitem[{{Polishook} {et~al.}(2017){Polishook}, {Jacobson}, {Morbidelli}, \&
  {Aharonson}}]{Polishook.et.al2017}
{Polishook}, D., {Jacobson}, S.~A., {Morbidelli}, A., \& {Aharonson}, O. 2017,
  Nature, 1, 0179

\bibitem[{{Popescu} {et~al.}(2012){Popescu}, {Birlan}, \& {Nedelcu}}]{M4AST}
{Popescu}, M., {Birlan}, M., \& {Nedelcu}, D.~A. 2012, A\&A, 544, A130

\bibitem[{{Pravec} {et~al.}(2005){Pravec}, {Harris}, \& {18
  co-authors}}]{Pravec.et.al2005}
{Pravec}, P., {Harris}, A.~W., \& {18 co-authors}. 2005, Icarus, 173, 108

\bibitem[{{Raymond} \& {Izidoro}(2017)}]{RaymondIzidoro2017}
{Raymond}, S.~N. \& {Izidoro}, A. 2017, Icarus, 297, 134

\bibitem[{{Rivkin} {et~al.}(2003){Rivkin}, {Binzel}, {Howell}, {Bus}, \&
  {Grier}}]{Rivkin.et.al2003}
{Rivkin}, A.~S., {Binzel}, R.~P., {Howell}, E.~S., {Bus}, S.~J., \& {Grier},
  J.~A. 2003, Icarus, 165, 349

\bibitem[{{Rivkin} {et~al.}(2002){Rivkin}, {Brown}, {Trilling}, {Bell}, \&
  {Plassmann}}]{Rivkin.et.al2002}
{Rivkin}, A.~S., {Brown}, R.~H., {Trilling}, D.~E., {Bell}, J.~F., \&
  {Plassmann}, J.~H. 2002, Icarus, 156, 64

\bibitem[{{Rivkin} \& {Emery}(2010)}]{RivkinEmery2010}
{Rivkin}, A.~S. \& {Emery}, J.~P. 2010, Nature, 464, 1322

\bibitem[{{Rivkin} {et~al.}(2007){Rivkin}, {Trilling}, {Thomas}, {DeMeo},
  {Spahr}, \& {Binzel}}]{Rivkin.et.al2007}
{Rivkin}, A.~S., {Trilling}, D.~E., {Thomas}, C.~A., {et~al.} 2007, Icarus,
  192, 434

\bibitem[{{Sanchez} {et~al.}(2014){Sanchez}, {Reddy}, {Kelley}, {Cloutis},
  {Bottke}, {Nesvorn\'{y}}, {Lucas}, {Hardersen}, {Gaffey}, {Abell}, \&
  {Corre}}]{Sanchez.et.al2014}
{Sanchez}, J.~A., {Reddy}, V., {Kelley}, M.~S., {et~al.} 2014, Icarus, 228, 288

\bibitem[{{Scholl} {et~al.}(2005){Scholl}, {Marzari}, \&
  {Tricarico}}]{Scholl.et.al2005}
{Scholl}, H., {Marzari}, F., \& {Tricarico}, P. 2005, Icarus, 175, 397

\bibitem[{{Spencer}(1990)}]{Spencer1990}
{Spencer}, J.~R. 1990, Icarus, 83, 27

\bibitem[{{Takir} \& {Emery}(2012)}]{TakirEmery2012}
{Takir}, D. \& {Emery}, J.~P. 2012, Icarus, 219, 641

\bibitem[{{Tholen}(1984)}]{TholenPhD}
{Tholen}, D.~J. 1984, PhD thesis, University of Arizona, Tucson

\bibitem[{{Vernazza} {et~al.}(2016){Vernazza}, {Marsset}, {Beck}, {Binzel},
  {Birlan}, \& {et al.}}]{Vernazza.et.al2016}
{Vernazza}, P., {Marsset}, M., {Beck}, P., {et~al.} 2016, Astron.~J., 152,
  id.~54

\bibitem[{{Vilas} \& {Gaffey}(1989)}]{VilasGaffey1989}
{Vilas}, F. \& {Gaffey}, M.~J. 1989, Science, 246, 790

\bibitem[{{Vilas} {et~al.}(1994){Vilas}, {Jarvis}, \&
  {Gaffey}}]{Vilas.et.al1994}
{Vilas}, F., {Jarvis}, K.~S., \& {Gaffey}, M.~J. 1994, Icarus, 109, 274

\bibitem[{{Vilas} {et~al.}(1998){Vilas}, {Smith}, {McFadden}, {Gaffey},
  {Larson}, {Hatch}, \& {Jarvis}}]{Vilas}
{Vilas}, F., {Smith}, B., {McFadden}, L., {et~al.} 1998, Vilas Asteroid Spectra
  V1.1, eAR-A-3-RDR-VILAS-ASTEROID-SPECTRA-V1.1. NASA Planetary Data System

\bibitem[{{Vokrouhlick{\'y}} {et~al.}(2016){Vokrouhlick{\'y}}, {Bottke}, \&
  {Nesvorn{\'y}}}]{Vokrouhlicky.et.al2016}
{Vokrouhlick{\'y}}, D., {Bottke}, W.~F., \& {Nesvorn{\'y}}, D. 2016, \aj, 152,
  39

\bibitem[{{Walsh} {et~al.}(2011){Walsh}, {Morbidelli}, {Raymond}, {O'Brien}, \&
  {Mandell}}]{Walsh.et.al2011}
{Walsh}, K.~J., {Morbidelli}, A., {Raymond}, S.~N., {O'Brien}, D.~P., \&
  {Mandell}, A.~M. 2011, Nature, 475, 206

\end{thebibliography}

\end{document}